\documentclass[conference]{IEEEtran}
\usepackage{hyperref}

\usepackage{multicol,multirow}

\ifCLASSINFOpdf 
\ifCLASSINFOpdf
\usepackage[pdftex]{graphicx}
\else
\fi
\begin{document}
\title{Dependency Update Adoption Patterns in the Maven Software Ecosystem}
\author{
    \IEEEauthorblockN{Baltasar Berretta, Augustus Thomas, Heather Guarnera}  
    \IEEEauthorblockA{Department of Mathematical and Computational Sciences, The College of Wooster, USA}  
    \IEEEauthorblockA{sberrettamagarinos25@wooster.edu, gthomas25@wooster.edu, hguarnera@wooster.edu}  
}

\maketitle

\begin{abstract}
Regular dependency updates protect dependent software components from upstream bugs, security vulnerabilities, and poor code quality. Measures of dependency updates across software ecosystems involve two key dimensions: the time span during which a release is being newly adopted (adoption lifespan) and the extent of adoption across the ecosystem (adoption reach). We examine correlations between adoption patterns in the Maven software ecosystem and two factors: the magnitude of code modifications (extent of modifications affecting the meaning or behavior of the code, henceforth called ``semantic change") in an upstream dependency and the relative maintenance rate of upstream packages. Using the Goblin Weaver framework, we find adoption latency in the Maven ecosystem follows a log-normal distribution while adoption reach exhibits an exponential decay distribution.

\end{abstract}


% \fi
\IEEEpeerreviewmaketitle

\section{Introduction}

Software ecosystems have become increasingly interconnected, with modern applications relying heavily on third-party libraries and frameworks~\cite{10.1145/3593434.3593474}. This interdependence, while fostering code reuse and accelerating development, introduces challenges in managing dependencies effectively. The Maven ecosystem, centered around the Apache Maven build automation tool, stands as a prime example of this complex web of dependencies in the Java development world.

Apache Maven has revolutionized Java project management by providing a standardized build system and dependency management tool. It simplifies the build process through a project object model (POM) and a set of plugins, offering features such as automated dependency resolution, uniform build lifecycles, and project information management \cite{yang2023towards}. The widespread adoption of Maven in the Java community has led to the creation of the Maven Central Repository, a vast collection of reusable Java libraries.

The Maven Central Graph offers researchers and practitioners a unique opportunity to study the dynamics of large-scale software ecosystems~\cite{jaime_goblin_2024}. This graph captures the intricate relationships between projects, revealing patterns of dependency usage, update behaviors, and potential vulnerabilities that can propagate through the ecosystem~\cite{wu2023understanding}. While the Maven ecosystem enables code reuse and modular development, it also presents unique development challenges such as dependency conflicts, maintenance overhead, and rapid dependency adoption.

Given these challenges, this work investigates the factors which correlate with adoption patterns in the Maven ecosystem. We consider adoption patterns for a given release as measured in two ways: the number of downstream dependents (i.e., popularity) and the varying speeds in which dependent packages adopt the new release. We define the range of the latter as a version's \textit{adoption lifespan}, that is, the difference between the latest adoption time and the earliest adoption time. We investigate adoption patterns as a function of two factors: semantic change size and maintenance activity. Specifically, we address the following research questions:

\textbf{RQ1: How does semantic change size correlate with adoption patterns?}  We examine correlations between the magnitude of changes (major, minor, or patch) in a dependency and its number of dependents and adoption lifespan. This builds on previous research on library migration patterns~\cite{he2021large}, extending it to the specific context of update latency. In addition, we anticipate highly depended-upon projects may face greater pressure or resistance to updates.

\textbf{RQ2: How does maintenance activity correlate with adoption patterns?}  We explore the relationship between a project's overall maintenance activity and its dependents' tendency to update promptly. This involves comparing measures of release frequency and adoption lifespan. 

By investigating these factors, we aim to provide insights that can help both library maintainers and consumers make informed decisions about dependency management strategies. Maintainers would benefit from considering the scope and speed at which consumers adopt their changes. Consumers might also benefit from understanding development time costs of dependency management. Understanding the dynamics of dependency updates can lead to improved tools, practices, and policies that enhance the overall health and security of the Maven ecosystem.

The main contribution of this study is an analysis of dependency update patterns in the Maven ecosystem, considering the impacts of dependents, semantic versioning, and maintenance activity. We find that the number of dependents has a positive relationship with the minimum adoption lifespan for releases, larger semantic changes had higher adoption lifespans, and highly maintained packages had lower adoption lifespans. We also found that releases corresponding to larger semantic changes had more dependents, as well as releases with low and medium maintenance rates.

This paper is organized as follows.
Section~\ref{sec: lit review} discusses related works and Section~\ref{sec: methodology} presents the methodology, including the data collection and relevant adoption pattern terminology. The analysis and discussion of influencing factors, including maintenance rate and semantic change, is given in Section~\ref{sec: analysis}. Section~\ref{sec: threats} details threats to validity, and Section~\ref{sec: conclusion} concludes.
%
%Section \ref{sec: lit review} discusses the state of the literature. Section~\ref{sec: adoption latency} summarizes adoption lifespan in the Maven ecosystem overall. Section~\ref{sec: dependent number} discusses adoption lifespan as a function of number of dependents. Section \ref{sec: size} discusses adoption latency as a function of semantic change size. Section \ref{sec: maint} discusses adoption latency as a function of maintenance rate. In Section \ref{sec: findings} we hope to offer actionable insights for improving the resilience and efficiency of interdependent software projects.

\section{Related Works} \label{sec: lit review}
Many researchers in the past have generated Maven Central Dependency Graph (DG) datasets using the Dependency Graph Mining Framework (DGMF)~\cite{MavenDependGraph,MavenRepoDataset,DGMF}. Some development tools for safe dependency update checking have been built using this mining framework~\cite{UpCy}. Recently, researchers have developed the Goblin framework for enriching and querying the Maven Central DG~\cite{jaime_goblin_2024}\cite{Goblin}.
%; it includes a DG meta-model, a miner, and an on-demand metrics weaver. The weaver API helps researchers easily build their own custom metrics for analysis of the Maven Central DG. 
These tools have been used to analyze the Maven Central DG in numerous ways, such as
characterizing vulnerability propagation through dependencies \cite{wu_yu_2013},
direct dependency update rhythms~\cite{jaime_goblin_2024},
general adoption trends \cite{kula_gaikovina_2018},
library maintenance trends,
and to distinguish between sustainable software and vulnerable software packages~\cite{MavenRepoDataset}. 

While these tools provide the infrastructure for analysis, understanding how to measure and interpret maintenance patterns requires examining established software maintenance metrics. Software maintenance rates are measured through quantitative metrics and qualitative assessments. Key maintenance metrics are resolved maintenance requests per time unit, mean time to repair (MTTR), and effort measured in person-days~\cite{sneed_measuring_1997}. Organizations track maintenance types (corrective, adaptive, perfective, preventive)~\cite{fasolino_metrics_2000} and measure maintenance impact through lines of code or function points modified~\cite{henry_defining_1996}. Quality assessment metrics include introduced defects and post-release defect density~\cite{fasolino_metrics_2000}. Maintenance costs as a portion of IT budget are also considered~\cite{sneed_measuring_1997}. Qualitative assessments, such as user satisfaction surveys, complement these metrics. Advanced analytics help identify trends and optimize resource allocation~\cite{henry_defining_1996}. These metrics provide insight into how maintenance rates might be measured as influencing dependency update latency in Maven.

\hbadness=99999
Studies employ various methods to quantify semantic change sizes in software evolution. Researchers categorize dependency constraints as compliant, permissive, or restrictive relative to semantic versioning rules~\cite{raemaekers_semantic_2017}. A more granular approach defines four semantic change relations: Modified Callsites, Modified Branch Conditions, New Value Propagation, and Modified Variables~\cite{zhang_has_2022}. This method uses abstract interpretation and AST differencing to measure change impact sets.
%It reduces false positives by 23-49 percent and change impact set sizes by 19-91 percent compared to conventional techniques \cite{zhang_has_2022}.
These methodologies offer insights into how change sizes might be measured as affecting dependency update latency in software ecosystems like Maven.

There are many approaches to analyze software dependencies in ecosystems. Sharma et al.~\cite{sharma_dependency_2009} use Java HashMaps to capture dependency information, introducing Incoming and Outgoing Interaction Density metrics. Zhang et al. \cite{zhang_feature-driven_2006} focus on feature-level dependencies, identifying refinement, constraint, influence, and interaction types. Robillard ~\cite{robillard_topology_2008} studies structural dependencies in source code, using fuzzy sets to rank relevant program elements for developers. While Sharma et al. and Zhang et al. emphasize component and feature-level analysis respectively, Robillard's method operates at program element level. These approaches improve dependency maintenance, contributing to our understanding of how dependencies might relate to adoption latency in the Maven ecosystem.

%We will study the time required to update a dependency from its latent state to its trusted state.
This study evaluates adoption patterns by number of dependents and adoption lifespan, as influenced by two important factors: semantic change size (RQ1) and maintenance rate (RQ2).

\section{Methodology} \label{sec: methodology}

\subsection{Dataset Construction}
To conduct this study, we leverage the Goblin framework, which provides tools for enriching and querying the Maven Central DG~\cite{jaime_goblin_2024}. We access the Maven Central DG through a Neo4J database infrastructure as described in~\cite{jaime_goblin_2024}. Our dataset contains detailed information about Maven packages, including their release dates, version histories, and dependency relationships, current as of November 25th, 2024.

The Neo4J database follows a graph representation in which nodes are Maven packages and edges are dependency relationships between packages. Each node contains metadata including the package's group ID, artifact ID, version information, and release timestamp. The edges contain information about the dependency relationship type (direct or transitive) and version constraints from the package's Project Object Model (POM) file. Through the Cypher query language, we extract and analyze this network of relationships to understand dependency update patterns.

To ensure data quality and relevance, we apply several filtering criteria. We exclude packages with incomplete metadata and test packages identified by standard Maven naming patterns. We focus on packages with at least one dependent and those following semantic versioning conventions. Our final dataset consists of approximately 7.5 million package versions across 380,000 unique artifacts, with over 30 million dependency relationships. For distribution analyses, we sample the 1,000 packages with the highest dependent numbers to characterize patterns in trusted packages.

\subsection{Adoption Lifespan}
We define adoption lifespan as the number of days between the first adoption of a dependency version and its last adoption by the community. This metric signifies the lifespan of a package version as implied by community adoption practices. Using Cypher queries, we calculate this by finding the temporal difference between earliest and latest dependency timestamps for each version.

\begin{figure}
    \centering
    \includegraphics[width=1.0\linewidth]{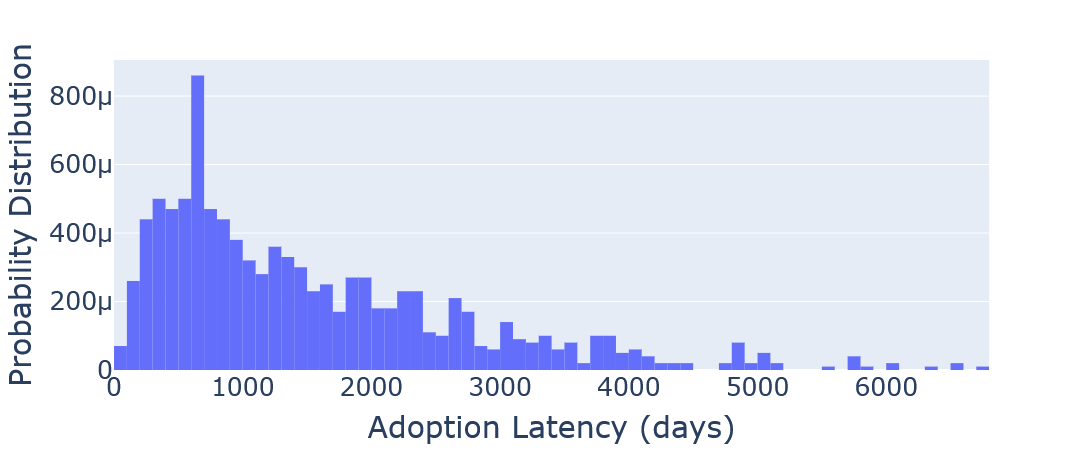}
    \caption{Probability distribution of adoption lifespan across a sample of Maven packages $\left(N=1,000\right)$}
    \label{fig:adoptionLatency}
\end{figure}

\subsection{Dependent Analysis}
We analyze dependency relationships through graph traversal queries identifying the number and nature of dependent connections. For each package version, we count unique downstream dependents and analyze their temporal adoption patterns. Our queries filter for direct dependencies where version constraints follow semantic versioning patterns.

\begin{figure}
    \centering
    \includegraphics[width=1.0\linewidth]{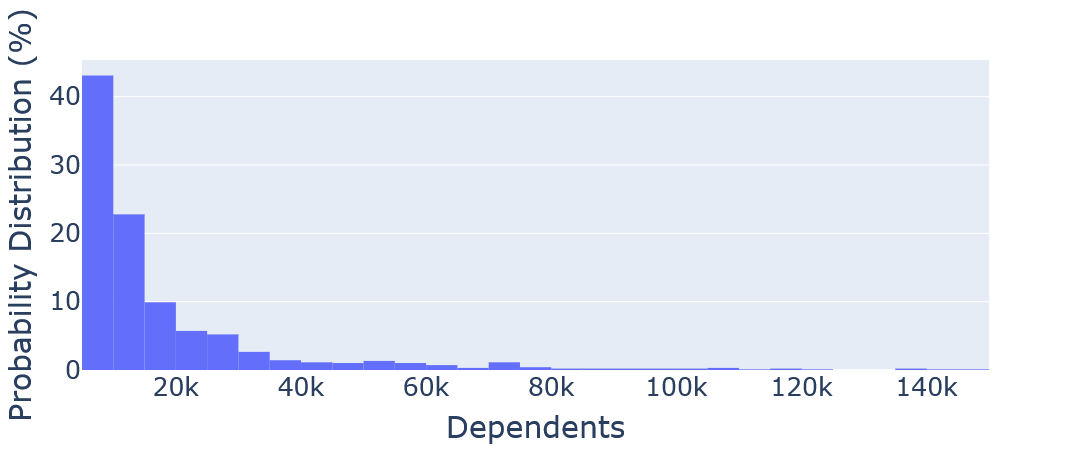}
    \caption{Probability distribution of number of dependents across a sample of Maven packages $\left(N=1,000\right)$}
    \label{fig:dependents}
\end{figure}

\begin{figure}
    \centering
    \includegraphics[width=1.0\linewidth]{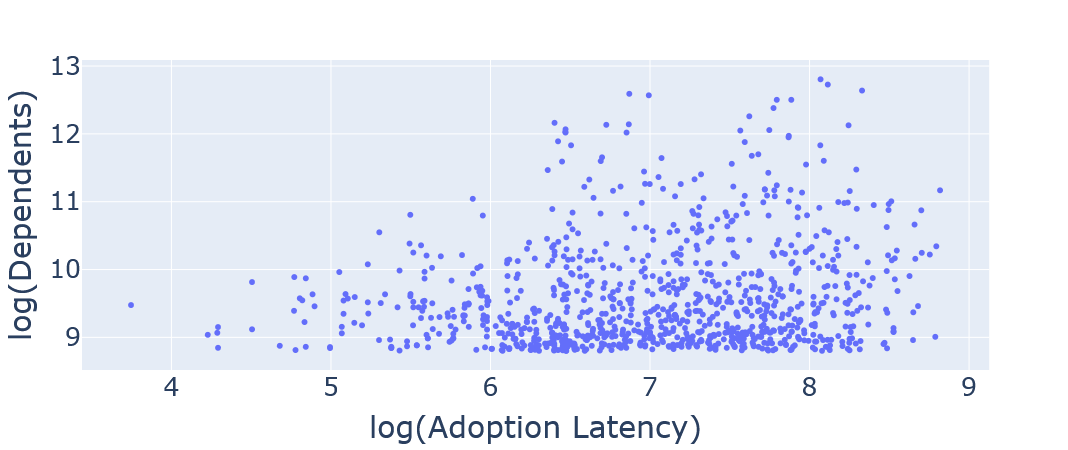}
    \caption{Logarithm of dependent number as a function of the logarithm of adoption lifespan}
    \label{fig:dependentsVersusLatency}
\end{figure}

A full replication package is available \footnote{Replication package available here: \href{https://zenodo.org/records/14291958?token=eyJhbGciOiJIUzUxMiJ9.eyJpZCI6IjQ2OTNkZTdkLWUwOGQtNGQwMC1iMzcwLWE4YjdiN2E2ZGI4YiIsImRhdGEiOnt9LCJyYW5kb20iOiJiYTc2OGIxNzM5ZjQ3YjdiOGE2NDVjZDA4ZTMwMGUwZCJ9.eVfudwlLzrDaqvnp972kIAFKRUVgwLUnmpAI3n9mAaHBk2Yy6eIdGCVwxzzBqTDDE1FqdVdYoIZuk5et19A9fA}{Zenodo repository}}.

\section{Results and Discussion} 
\label{sec: analysis}
We now consider the adoption lifespan and the dependent number adoption patterns according to two different parameters - semantic change size and maintenance activity - which are formally described next.

\subsection{General Distribution Patterns}

Fig.~\ref{fig:adoptionLatency} shows the distribution of adoption lifespans in the Maven ecosystem for releases with the top number of dependents (N=1,000). The distribution follows a logarithmic normal distribution curve $\left(\mu = 7.05, \sigma = 0.785\right)$. 
Log-normal distribution suggests that the largest packages in the ecosystem vary widely in their adoption timelines. The modal number of days between the first and last adoption of a package is $625$ days, which is also an outlier. Therefore, a statistically significant subset of packages are effectively going end-of-life and unadopted after exactly two years in the ecosystem. 

Fig.~\ref{fig:dependents} shows the distribution of the number of dependents in the Maven ecosystem. Our sample follows a stark exponential decay distribution. $99\%$ of packages have between 6.8k and 80k dependents, with the median package having 17k dependents. These dependent distributions indicate a highly clustered and sparsely connected dependency network, in which a small fraction of packages have dependent numbers which are orders of magnitude higher than the mode.

There is a positive relationship between number of dependents and the lower bound on adoption lifespan, as can be seen in Fig.~\ref{fig:dependentsVersusLatency}; as a package is used by more dependents, the likelihood that it will be newly adopted increases. This suggests a relationship between longevity and number of dependents. However, there is no correlation in general between the logarithm of these two variables: the Spearman correlation is 0.16 and the Kendall correlation is 0.11.

\subsection{Semantic Change Size (RQ1)} \label{sec: size}
% TODO: verifying -- this does not assume #1? That is, 2.0.0 to 3.0.0 is a major change still?
We define semantic change size following the conventions of semantic versioning, in which version numbers are separated into three parts: ``major.minor.patch''. For example, a version change from ``1.X.X" to ``2.0.0" is a major change, whereas ``X.1.X" to ``X.2.0" is a minor change. A version change from ``X.X.1" to ``X.X.2" is a patch. If maintainers adhere to semantic versioning conventions, release size can be read directly from version number. For example, ``1.2.3" indicates a patch release as it changes only the rightmost number. ``1.1.0" indicates a minor release as shown by a change in the middle number with a zero patch number. Finally, ``2.0.0" indicates a major release as shown by the change in the leftmost number with zeroed minor and patch numbers. For the purposes of observability in this paper, we assume that Maven developers adhere well to semantic versioning conventions. 

In general, major releases imply breaking changes, are not backwards compatible, and are therefore most difficult to adopt; downstream dependents must refactor and test before adoption is successful. Minor releases add backwards compatible functionality; downstream dependents can adopt new features without significant development costs. Patches are minor bug fixes and optimizations which should not require downstream reworks.

% TODO: what if x = 0.1 ?
\begin{table*}[ht]
\caption{Adoption factors according to varying semantic change size and maintenance rate.}
\centering
\begin{tabular}{|c|ccc|ccc|}
\hline
\multirow{2}{*}{} & \multicolumn{3}{c|}{\textbf{Semantic Change Size}} & \multicolumn{3}{c|}{\textbf{Maintenance Rate}} \\ \cline{2-7} 
                                            & \textbf{Major}     & \textbf{Minor}    & \textbf{Patch}   & \textbf{High}      & \textbf{Medium}    & \textbf{Low}      \\ 
                                            & \textbf{C}.X.X    & X.\textbf{C}.X      & X.X.\textbf{C}   & $x > 0.1$      & $0.01 \le x \le 0.1$      & $x < 0.01$      \\ \hline
\textbf{Number of packages}                 & 3,526,059 (84\%)           & 587,676 (14\%)          & 83,954 (2\%)          & 1,830,042 (44\%)    & 181,600 (4\%)      & 2,186,047 (52\%)   \\ \hline
\textbf{Average dependents}                 & 17.65              & 8.09              & 4.21             & 8.58               & 19.73              & 22.05              \\ \hline
\textbf{Average adoption lifespan (in days)} & 34.52             & 25.78             & 19.73            & 0.36               & 39.88              & 59.84               \\ \hline
\end{tabular}

\label{table:semantic_maintenance}
\end{table*}

The distribution of semantic change sizes across packages in the Maven ecosystem, along with the adoption patterns of each category, is summarized in the first half of Table~\ref{table:semantic_maintenance}.
%\begin{figure}
%    \centering
%    \includegraphics[width=1.0\linewidth]{SemanticChangeSizeDist.png}
%    \caption{Histogram showing the distribution of major, minor, and patch change releases}
%    \label{fig:SemanticChangeSizeDistribution}
%\end{figure}
%Fig.~\ref{fig:SemanticChangeSizeDistribution} shows the semantic changes sizes across packages in the Maven ecosystem. 
The vast majority of package releases correspond to major changes, being 84\% of all releases. A small minority are minor changes, around 14\%. According to our method, almost no releases are patches. This suggests package developers prefer major and minor releases over constant patches.
%\begin{figure}
%    \centering
%    \includegraphics[width=1.0\linewidth]{semanticAdoptionLatency.png}
%    \caption{Average adoption latency of major, minor, and patch change releases}
%    \label{fig:semanticAdoption}
%\end{figure}
%Fig.~\ref{fig:semanticAdoption} shows the average days to adoption as a function of semantic change size.
Patch, minor, and major version changes take an average of five, six, and seven weeks to adopt respectively. As expected, larger semantic differences make adoption more difficult.
%\begin{figure}
%    \centering
%    \includegraphics[width=1.0\linewidth]{semanticDependents.png}
%    \caption{Average dependents of major, minor, and patch change releases}
%    \label{fig:semanticDependents}
%\end{figure}
%Fig.~\ref{fig:semanticDependents} shows the average number of dependents as a function of semantic change size. 
Our analysis shows that major releases positively correlate with higher numbers of dependents, followed by minor then patch releases. The data suggests that software developers in the Maven ecosystem tend to have longer adoption times for major releases. Additionally, dependent maintainers may exclusively track major versions rather than continuously adopting less important minor or patch changes. 

\subsection{Maintenance Rate (RQ2)} \label{sec: maint}
We define a package’s maintenance rate as its number of unique versions used per year, calculated as (unique versions)/(adoption span in years). This metric quantifies how frequently a package changes relative to how often those changes are adopted. If a package has no adoption span, its maintenance rate equals the number of unique versions. We also define the maintenance rate ratio as the average number of new version releases per adoption lifespan window. Packages are categorized based on maintenance rate $x$: high ($x > 0.1$), medium ($0.01 < x < 0.1$), and low ($x < 0.01$).

% TODO: insert a high vs. low example for clarity

%Table \ref{tab:maintanence} shows our definitions of high, medium, and low maintenance rates. (see Section \ref{sec: adoption latency} for adoption latency). 
%\begin{table}[]
%    \centering
%    \begin{tabular}{c|c}
%        \hline
%         Maintenance Rate & Ratio \\
%         High & $x > 0.1$\\
%         Medium & $0.01 < x < 0.1 $\\
%         Low & $x < 0.01$\\
%         \hline
%    \end{tabular}
%    \caption{Maintenance Rate Metric Definition}
%    \label{tab:maintanence}
%\end{table}

%\subsection{Findings}

%\begin{figure}
%    \centering
%    \includegraphics[width=1.0\linewidth]{maintenanceNumber.png}
%    \caption{Histogram showing the distribution of high, medium, and low maintenance rate packages}
%    \label{fig:maintenanceNumber}
%\end{figure}
The distribution of packages with high, medium, and low maintenance rates, along with the adoption patterns of each category, is summarized in the second half of Table~\ref{table:semantic_maintenance}.
%
%Fig.~\ref{fig:maintenanceNumber} shows the distribution of packages with high, medium, and low maintenance levels. 
Most projects correspond to high or low maintenance rates. This suggests a polarizing view on maintenance scheduling across the Maven ecosystem. Our maintenance rate metric also measures the community's simultaneous adoption of multiple versions of a given project. This distribution therefore also suggests that some sub-ecosystems in Maven support far fewer simultaneous versions than others. 

%\begin{figure}
%    \centering
%    \includegraphics[width=1.0\linewidth]{adoptionMaintence.png}
%    \caption{Average adoption latency of high, medium, and low maintenance rate packages}
%    \label{fig:adoptionMaintenance}
%\end{figure}

%Fig.~\ref{fig:adoptionMaintenance} shows the average adoption latencies of packages with high, medium, and low maintenance rates. As expected, highly maintained packages have faster adoption speeds, whereas low maintenance packages are adopted over longer periods of time.
%\begin{figure}
%    \centering
%    \includegraphics[width=1.0\linewidth]{dependentMaintence.png}
%    \caption{Average dependents of high, medium, and low maintenance rate packages}
%    \label{fig:dependentMaintenance}
%\end{figure}

%Fig.~\ref{fig:dependentMaintenance} shows the average dependents per release for packages with high, medium, and low maintenance rates. 
As expected, packages with high maintenance rates tend to have lower dependents per release. Dependents may not see minor releases as more adoptable than patch releases. Dependents tend to adopt a wide range of different releases when their dependencies are frequently updated. This finding agrees with our semantic change size findings; patches and quickly updated versions are adopted less frequently than their major and slowly updated counterparts. %\\

\section{Threats to validity} \label{sec: threats}

The most significant threat to validity in this study is the dataset sampling methods. For our analysis of the distributions of adoption lifespan and adoption reach across Maven, we only investigate the 1,000 packages with the highest number of dependents; we sample due to limitations on infrastructure and query time. While this helps us understand patterns in widely-used packages, it limits generalizability to the broader ecosystem. Less popular packages may exhibit different adoption patterns due to factors such as smaller user bases, different maintenance practices, or specialized use cases.

A second threat is our assumption about semantic versioning compliance. Although semantic versioning conventions are widely adopted, developers may not consistently follow these guidelines. Version numbers may not accurately reflect the magnitude of changes, and what constitutes a "breaking change" can be subjective. This limitation could affect our analysis of how semantic change size relates to adoption patterns. However, it is the opinion of the authors that version number remains a reasonably valid measure for semantic change sizes in package releases. 

Our analysis also does not fully capture the complexity of ecosystem dynamics and developer behavior. Factors such as organizational policies, automated dependency updates, security considerations, and developer preferences could influence adoption patterns in ways our metrics cannot measure. Community dynamics, like the relationship between maintainers and users, or the impact of documentation quality, are also not captured in our quantitative analysis.

These limitations suggest opportunities for future research combining our quantitative findings with qualitative studies of developer behavior and ecosystem dynamics.

\section{Conclusion} \label{sec: conclusion}
We examined the relationships between semantic change size, maintenance rate, and package adoption patterns in the Maven software ecosystem. We found that dependent number has a positive relationship on the minimum adoption lifespan for a package. Our analysis shows that larger semantic changes correlate with higher adoption lifespans. The data indicates that highly maintained packages are associated with lower adoption lifespans.

We also investigated the effect of semantic change size and maintenance rate on package dependent number. Our analysis reveals that releases with larger semantic changes are associated with higher numbers of dependents. The data shows that releases with low and medium maintenance rates correlate with higher numbers of dependents. 

These findings suggest several practical strategies for both maintainers and consumers. Maintainers of highly-depended packages should expect longer adoption periods for major releases, and might consider providing detailed migration guides. Consumers should allocate more time for adopting major version changes, especially from packages with many dependents. For packages with high maintenance rates, automated dependency update tools may be particularly valuable given the shorter adoption windows observed. A number of open questions remain for future work.
One avenue is to disentangle the relationships between release rate and semantic versioning. In general, patches are released more frequently than major and minor versions. One could further granulate the differences in adoption latency between different maintenance rates of the same semantic change type.

Future work can also investigate tendencies to adopt minor changes. This work found that minor changes and patches have nearly the same adoption lifespan. We infer that development teams in the Maven ecosystem are pulling new features in minor changes with equal frequency as they pull patches, without requiring the added functionality. Future researchers might investigate whether this tendency has adverse security risks, or otherwise effects the Maven ecosystem. 

Future work might explore the use of adoption lifespans across the Maven ecosystem to quantify semantic changes in source code. As we have seen, larger semantic version changes correlate to longer adoption lifespans. Researchers aiming to benchmark static analysis for semantic change estimation might use adoption lifespan as a proxy measure.

This work began investigation on release lifespans. Future work might characterize the overlap between the lifespans of simultaneously supported and adopted versions of the same dependency in the community. Researchers might investigate how the Maven ecosystem tends to concentrate or disperse their adoption across multiple versions of the same package. Developers might be interested in release scheduling which effectively concentrates adoption patterns onto stable versions.

\bibliographystyle{IEEEtran}
\bibliography{bib/paper.bib}

\end{document}